\documentclass[aps,preprint]{revtex4}
\usepackage{amsfonts}
\usepackage{amsmath}
\usepackage{amssymb}
\usepackage{graphicx}

\newtheorem{theorem}{Theorem}
\newtheorem{acknowledgement}[theorem]{Acknowledgement}

\input{tcilatex}

\begin{document}

\preprint{}
\title[ ]{Black holes and the classical model of a particle in Einstein
non-linear electrodynamics theory}
\author{S. Habib Mazharimousavi}
\email{habib.mazhari@emu.edu.tr}
\affiliation{Department of Physics, Eastern Mediterranean University, G. Magusa, north
Cyprus, Mersin 10, Turkey.}
\author{M. Halilsoy}
\email{mustafa.halilsoy@emu.edu.tr}
\affiliation{Department of Physics, Eastern Mediterranean University, G. Magusa, north
Cyprus, Mersin 10, Turkey.}
\pacs{PACS number}

\begin{abstract}
Modified by a logarithmic term, the non-linear electrodynamics (NED) model
of the Born-Infeld (BI) action is reconsidered. Unlike the standard BI
action, this choice provides interesting integrals of the Einstein-NED
equations. It is found that the spherical matching process for a regular
black hole entails indispensable surface stresses that vanish only for a
specific value of the BI parameter. This solution represents a classical
model of an elementary particle whose radius coincides with the horizon. In
flat space time, a charged particle becomes a conducting shell with a radius
proportional to the BI parameter.
\end{abstract}

\maketitle

\section{Introduction}

Non-linear Born-Infeld (BI) theory was introduced to resolve the Coulomb
divergences of classical electrodynamics \cite{1}. With the advent of
quantum electrodynamics, it was all but forgotten until its reemergence
within the context of string theory. However, the original BI theory was
later extended to cover more general non-linear electrodynamics (NED)
theories\cite{2}. The NED action, with its square root term restricted to
real values, provides a natural way to avoid the Coulomb field's
singularity. This is reminiscent of the relativistic particle Lagrangian
that restricts the speed of a particle to less than the speed of light.

It was expected that the therapeutic effect of the BI action played a
non-trivial role when coupled with other fields. Gravity is no exception,
and a search for regular black hole solutions of the full theory attracted
much interest \cite{3}. Specifically, the existence of regular, purely
electrically-charged black holes continued to be a source of discussion\cite%
{4}. Within the context of the full Einstein-Yang-Mills-Born-Infeld theory
it was shown that regular magnetic black holes are a reality, while the pure
electrical ones remained on questionable footing \cite{5}. Our present
results use a new method that suggests the latter class, although not
generic, are quite real as well. Past efforts to study NED introduced a dual
structure, through a Legendre transformation, in which the NED solutions
were readily available.

In this Letter, without invoking any dual structures, we extend the BI
action by a novel non-polynomial term that admits regular black holes. In
the absence of gravity, it is clear that our NED model describes a charged
elementary particle of finite field energy with a natural cut-off, which
turns out to be the radius of the particle. This corresponds to the
classical glue-balls of Yang-Mills (YM) theory\cite{5} with the important
difference that the non-linear YM field is replaced here by the NED\cite{6}.
We concern ourselves entirely with spherically symmetric NED. We glue two
spacetimes together in such a manner that continuity of metric and certain
first derivatives are satisfied. As could be expected, this imposes severe
restrictions on the component metrics and the BI electric field. It is
possible, however, with the choice of a Bertotti-Robinson ($BR$) type ($%
\widetilde{BR}$) metric for interior and a Reissner-Nordstrom ($RN$) type ($%
\widetilde{RN}$) metric for the exterior\cite{7,8}. With a particular choice
of the BI parameter, it is shown both from the time-like and the null-shell
formalisms that the surface stress energy tensor, i.e., the Lanczos tensor, $%
S_{\nu }^{\mu }=0.$ Intriguingly, this corresponds to a case where the
matching surface coincides with the double horizon of a regular black hole.

We organize the Letter as follows. In Sec. II, we consider NED in a flat
spacetime. Sec. III covers the gluing BR type and RN type spacetimes,
resulting in a regular solution. We conclude with a discussion of
interpretation in Sec. IV.

\section{NED in Flat Spacetime}

With unit conventions assumed such that $(c=\hslash =k_{B}=8\pi G=\frac{1}{%
4\pi \epsilon _{\circ }}=1)$ our action $S$ and line element are 
\begin{eqnarray}
S &=&-\frac{1}{2}\int d^{4}x\sqrt{-g}\mathcal{L}\left( F,^{\star }F\right) ,
\\
ds^{2} &=&-dt^{2}+dr^{2}+r^{2}d\Omega ^{2},
\end{eqnarray}%
where%
\begin{equation}
d\Omega ^{2}=d\theta ^{2}+\sin ^{2}\theta d\varphi ^{2},
\end{equation}%
and 
\begin{equation}
\mathcal{L}=-\frac{2}{b^{2}}\left\{ 1-\sqrt{1+2b^{2}F-b^{4}\left( {}^{\star
}F\right) ^{2}}+\ln \left[ \frac{1}{2}\left( 1+\sqrt{1+2b^{2}F-b^{4}\left(
{}^{\star }F\right) ^{2}}\right) \right] \right\}
\end{equation}%
in which $b$ is the BI parameter, $F=F_{\mu \nu }F^{\mu \nu },$ $^{\star
}F=F_{\mu \nu }{}^{\star }F^{\mu \nu }$ and $^{\star }$ stands for duality.
Since we shall confine ourselves entirely to the electrostatic problem the $%
F_{\mu \nu }{}^{\star }F^{\mu \nu }$ term under the square root vanishes and
will be ignored in the subsequent sections. The parameter $b$ is such that%
\begin{eqnarray}
\lim_{b\rightarrow 0}\mathcal{L} &=&F_{\mu \nu }F^{\mu \nu },\text{ \ \
(Maxwell case)} \\
\lim_{b\rightarrow \infty }\mathcal{L} &=&0\text{, \ \ (zero action)}.
\end{eqnarray}

The electric field 2-form with the radial electric field $E\left( r\right) $
is chosen as%
\begin{equation}
\mathbf{F}=E\left( r\right) dt\wedge dr
\end{equation}%
which leads to $F=F_{\mu \nu }F^{\mu \nu }=-2E\left( r\right) ^{2}$. This
must satisfy the NED equation 
\begin{equation}
d\left( \mathcal{L}_{F}\ {}^{\star }\mathbf{F}\right) =0
\end{equation}%
where ${}^{\star }\mathbf{F=}E\left( r\right) \ r^{2}\sin \theta \ d\theta
\wedge d\varphi $. Integrating the latter equation and considering the line
element (2) one finds%
\begin{equation}
\frac{\sqrt{-2F}}{1+\sqrt{1+2b^{2}F}}=\frac{C}{r^{2}}
\end{equation}%
where $C\in 
\mathbb{R}
^{+}$ is a constant of integration. It is not difficult to show that this
equation gives a non-trivial solution%
\begin{equation}
F=\frac{-2C^{2}r^{4}}{\left( C^{2}b^{2}+r^{4}\right) ^{2}},
\end{equation}%
which upon substitution into the equation (9) implies%
\begin{equation}
\frac{2Cr^{2}}{\left( C^{2}b^{2}+r^{4}\right) +\left\vert
C^{2}b^{2}-r^{4}\right\vert }=\frac{C}{r^{2}}
\end{equation}%
which is valid only for $r>\sqrt{Cb}$ if $C\neq 0$. This solution
corresponds to the electric field%
\begin{equation}
E\left( r\right) =\frac{Cr^{2}}{\left( C^{2}b^{2}+r^{4}\right) }
\end{equation}%
which after using the Maxwell limit 
\begin{equation}
\lim_{b\rightarrow 0}E=\frac{C}{r^{2}}
\end{equation}%
suggests identifying the constant $C$ as the charge of the particle, i.e., $%
C=q$. To find the charge distribution one may look at the region $r<r_{\circ
}$ ($r_{\circ }=\sqrt{qb}$), where the only possible solution of (8) under
the spherically symmetric flat spacetime and spherically symmetric electric
field corresponds to C=0, or equivalently, a zero electric field. Note that
the existence of the absolute value in (11), which arises from the square
root term, makes this choice indispensable. That is, $\left\vert
C^{2}b^{2}-r^{4}\right\vert =C^{2}b^{2}-r^{4}$ for $r^{4}<C^{2}b^{2}$ ($%
=r^{4}-C^{2}b^{2}$ for $r^{4}>C^{2}b^{2}$). When this is employed in (11),
for the consistency of the solution, we must choose $C=0$, leading
automatically to $E(r)=0$ for $r^{4}<C^{2}b^{2}.$ Whenever $C\neq 0,$ on the
other hand, (12) becomes the only acceptable solution for $r^{4}>C^{2}b^{2}.$
These results lead to a surface charge distribution of the particle of $\rho
=\frac{\delta \left( r-r_{\circ }\right) }{4\pi r_{\circ }^{2}}$ in which $%
\delta \left( r-r_{\circ }\right) $ denotes the Dirac delta function.
Consequently, one can easily show that the electric potential of the
particle is a constant value inside ($r<r_{\circ }$) and 
\begin{equation}
\phi \left( r\right) =\frac{\sqrt{2}q}{4r_{\circ }}\left[ \tanh ^{-1}\left( 
\frac{\sqrt{2}rr_{\circ }}{r^{2}+r_{\circ }^{2}}\right) +\tan ^{-1}\left( 
\frac{\sqrt{2}rr_{\circ }}{r^{2}-r_{\circ }^{2}}\right) \right]
\end{equation}%
for the outside ($r>r_{\circ }$) region. For $b\rightarrow 0,$ we recover
the Coulomb field for a charge located at $r=0$, and $r=r_{\circ }\neq 0$
provides a natural cut-off for the particle. The total energy density is $u=%
\frac{1}{2}\mathbf{E.D,}$ ($\mathbf{D}=\epsilon \mathbf{E,}$ with $\epsilon =%
\frac{\partial \mathcal{L}}{\partial F}=1+\left( \frac{r_{\circ }}{r}\right)
^{4}$) with total energy%
\begin{equation}
U=4\pi \int_{r_{\circ }}^{\infty }u\left( r\right) \ r^{2}dr=5.45\frac{q^{2}%
}{r_{\circ }}.
\end{equation}%
This amounts to a hard-core particle with charge density $\rho .$
Identifying $U=M$, $r_{\circ }$ is determined from the energy of the
particle. If $\frac{q^{2}}{2M}$ is identified as the classical
electromagnetic radius, $r_{e}$, then $r_{\circ }=10.90r_{e}.$

\section{REGULAR ELECTRIC\ BLACK\ HOLES\ IN\ EINSTEIN-NED\ THEORY}

In this section, a composite spacetime will be established consisting of a
region $(r\leqslant r_{\circ })$ of uniform electric field glued at $%
r=r_{\circ }$ to an outside region $(r>r_{\circ }).$ The proper junction
condition will dictate that $r_{\circ }$ must coincide with the horizon of
the entire spacetime. For this purpose, we choose our action as 
\begin{equation}
S=\frac{1}{2}\int d^{4}x\sqrt{-g}\left[ R-\mathcal{L}\left( F\right) \right]
,
\end{equation}%
in which $R$ is the Ricci scalar, and the given Lagrangian (4) is free of
magnetic fields. The Einstein-NED equation is%
\begin{equation}
G_{\mu }^{\ \nu }=T_{\mu }^{\ \nu }=-\frac{1}{2}\left[ \mathcal{L}\left(
F\right) \delta _{\mu }^{\ \nu }-4\mathcal{L}_{F}\left( F\right) F_{\mu
\lambda }F^{\nu \lambda }\right]
\end{equation}%
in which the electromagnetic field 2-form (7) satisfies the NED equation
(8). The static, spherically symmetric spacetimes satisfying the foregoing
equations and being glued at $r=r_{\circ }$ are%
\begin{eqnarray}
ds^{2} &=&-\tilde{f}\left( r\right) dt^{2}+\frac{dr^{2}}{\tilde{f}\left(
r\right) }+r_{\circ }^{2}d\Omega ^{2},\text{ \ \ }(r\leqslant r_{\circ }), \\
ds^{2} &=&-f\left( r\right) dt^{2}+\frac{dr^{2}}{f\left( r\right) }%
+r^{2}d\Omega ^{2},\text{ \ \ }(r>r_{\circ }).
\end{eqnarray}%
The choice of these metrics can be traced back to the form of the
stress-energy tensor (17), which satisfies $T_{0}^{0}-T_{1}^{1}=0$ and
consequently $G_{0}^{0}-G_{1}^{1}=0$, whose explicit form, on integration,
gives $\left\vert g_{00}g_{11}\right\vert =C=$constant. We need only choose
the time scale at infinity to make this constant equal to unity.

Nevertheless, for a spherically symmetric charge in EM theory the external
solution is known uniquely to be the RN metric. Therefore, to recover the RN
metric in the Maxwell limit $(b\rightarrow 0)$, we must consider an RN type
metric ansatz for $r>r_{\circ }.$ Further, since the outer RN metric was
glued consistently with the inner BR metric \cite{7}, it is natural to seek
a similar ansatz in the present problem as well. On the hypersurface $%
r=r_{\circ }$, the continuity of metrics is assumed, whereas some metric
derivatives are allowed to be discontinuous to allow for physical sources.

The field equations combined with the junction conditions will determine the
metric functions $f\left( r\right) ,$ $\widetilde{f}\left( r\right) $ and
the electric field $E\left( r\right) $. We note from NED equation (8) that
the electric field is uniform in the region $r<r_{\circ }$. Our solution can
be summarized as follows%
\begin{gather}
E\left( r\right) =\left\{ 
\begin{array}{cc}
\frac{q}{2r_{\circ }^{2}}, & r\leq r_{\circ } \\ 
\frac{qr^{2}}{r^{4}+r_{\circ }^{4}}, & r>r_{\circ }%
\end{array}%
\right. \\
f\left( r\right) =1-\frac{2M}{r}+\frac{q^{2}}{3r_{\circ }^{4}}r^{2}\ln
\left( \frac{r^{4}}{r^{4}+r_{\circ }^{4}}\right) + \\
\frac{q^{2}\sqrt{2}}{3rr_{\circ }}\tan ^{-1}\left( \frac{\sqrt{2}rr_{\circ }%
}{r^{2}-r_{\circ }^{2}}\right) -\frac{q^{2}\sqrt{2}}{6rr_{\circ }}\ln \left[ 
\frac{r^{2}+r_{\circ }^{2}-\sqrt{2}rr_{\circ }}{r^{2}+r_{\circ }^{2}+\sqrt{2}%
rr_{\circ }}\right] ,\text{ \ \ }r>r_{\circ }  \notag \\
\widetilde{f}\left( r\right) =C_{0}r^{2}+C_{1}r+C_{2}\ ,\text{ \ \ }r\leq
r_{\circ }
\end{gather}%
where 
\begin{eqnarray}
C_{0} &=&\frac{q^{2}}{r_{\circ }^{4}}\left( 1-\ln 2\right) ,  \notag \\
C_{1} &=&\frac{2M}{r_{\circ }^{2}}+\frac{q^{2}}{6r_{\circ }^{3}}\left( 8\ln
2-12-\sqrt{2}\left( \pi -\ln \left( 3-2\sqrt{2}\right) \right) \right) , \\
C_{2} &=&1-\frac{4M}{r_{\circ }}-\frac{q^{2}}{3r_{\circ }^{2}}\left( 2\ln
2-3-\sqrt{2}\left( \pi -\ln \left( 3-2\sqrt{2}\right) \right) \right) , 
\notag
\end{eqnarray}%
\bigskip in which $M$ is a mass related constant, $q=$charge and $r_{\circ }=%
\sqrt{bq}.$ Furthermore, as a result of satisfying the field equations, $%
r_{\circ }$ and $q$ are constrained by the condition%
\begin{equation}
r_{\circ }=q\sqrt{\ln 2}.
\end{equation}%
At this stage it is important to state that the metric function $f\left(
r\right) $ and $\widetilde{f}\left( r\right) $ satisfy\bigskip 
\begin{equation}
f\left( r_{\circ }\right) =\tilde{f}\left( r_{\circ }\right) =0.
\end{equation}

These conditions eliminate the possibility of quasi-black holes (QBH), which
are defined as objects on the verge of being extremal black holes. In order
to create such QBH, we would have to consider matching conditions $f\left(
r_{0}\right) =\widetilde{f}\left( r_{0}\right) \neq 0.$ These will not be
our concern here. The conditions (25) dictate that%
\begin{eqnarray}
\frac{M}{r_{\circ }} &=&\frac{1}{2}+\frac{1}{12\ln 2}\left[ \sqrt{2}\left(
\pi -\ln \left( 3-2\sqrt{2}\right) \right) -2\ln 2\right]  \notag \\
C_{1} &=&\frac{2}{r_{\circ }}\left( 1-\frac{1}{\ln 2}\right) , \\
C_{2} &=&\frac{1}{\ln 2}-1,  \notag
\end{eqnarray}%
which casts $\widetilde{f}\left( r\right) $ into%
\begin{equation}
\tilde{f}\left( r\right) =\left( r-r_{\circ }\right) ^{2}
\end{equation}%
for the specific choice%
\begin{equation}
r_{\circ }^{2}=\frac{1}{\ln 2}-1.
\end{equation}%
Let us add that a combination of (24), (28) and $r_{\circ }=\sqrt{bq}$
determines the value of the BI parameter as $b=0.55$. The finite scalar
invariants, such as the Ricci and Kretschmann of the line element (18), for $%
r<r_{\circ }$ are given respectively by%
\begin{eqnarray}
R &=&\frac{2}{r_{\circ }^{2}}\left( 1+r_{\circ }^{2}\right) , \\
K &=&\frac{4}{r_{\circ }^{4}}\left( 1+r_{\circ }^{4}\right) .  \notag
\end{eqnarray}%
Letting now%
\begin{equation}
r_{\circ }-r=\frac{1}{\bar{r}},\text{ \ \ }t=\bar{t},\text{ }
\end{equation}%
transforms the metric into%
\begin{equation}
ds^{2}=\frac{-d\bar{t}^{2}+d\bar{r}^{2}}{\bar{r}^{2}}+r_{\circ }^{2}d\Omega
^{2},\text{ \ \ }\left( r\leq r_{\circ }\right) .
\end{equation}%
This is a Bertotti-Robinson ($BR$)\cite{9,10} type metric with a specific
radius that will be referred to here as the $\widetilde{BR}$ spacetime.
Similarly we label the metric (19) for $r>r_{\circ },$ as the $\widetilde{RN}%
.$ It is well-known that the $BR$ metric is not a black hole solution.
However, our present $\widetilde{BR}$ is a part of a composite system of
spacetimes, with an event horizon at $r_{\circ }$, where it corresponds to
an accelerated frame in a conformally flat background with a unit
acceleration in the present context\cite{11}. Let us note that our result of 
$\widetilde{BR}$ \ for $\left( r<r_{\circ }\right) $ is not contradicted by
a theorem proved long ago by Bronnikov and Shikin\cite{12}. This theorem
proved the non-existence of a regular center, which is still satisfied in
the case of our $\widetilde{BR}$ spacetime in the Einstein-NED theory.

In order to determine if our matching of inner $\widetilde{BR}$ to outer $%
\widetilde{RN}$ is smooth, we compute the surface stress tensor $S_{\mu
}^{\nu }$ on $r=r_{\circ }$. This can be expressed in terms of the extrinsic
curvature tensor in accordance with 
\begin{equation}
8\pi S_{\mu }^{\nu }=[K_{\mu }^{\nu }]-\delta _{\mu }^{\nu }[K],
\end{equation}%
where $[.]=\left( .\right) _{+}-\left( .\right) _{-},$ with $K=K_{\mu }^{\mu
}$ and $\mu ,\nu =\left\{ t,\theta ,\varphi \right\} .$ Here $\left(
.\right) _{+}$ $\ $and $\left( .\right) _{-}$ refer to the outer $\left(
r>r_{\circ }\right) $ and the inner$\left( r<r_{\circ }\right) $ metrics,
respectively. The components of $S_{\mu }^{\nu }$ are given by\cite{7} 
\begin{eqnarray}
8\pi S_{0}^{0} &=&\frac{2}{r}\left[ \left( r_{+}\right) ^{\prime }-\left(
r_{-}\right) ^{\prime }\right] , \\
8\pi S_{2}^{2} &=&8\pi S_{3}^{3}=\frac{\left( r\sqrt{f\left( r\right) }%
\right) _{+}^{\prime }}{r\sqrt{f\left( r\right) }}-\frac{\left( r-r_{\circ
}\right) _{-}^{\prime }}{\left( r-r_{\circ }\right) },
\end{eqnarray}%
where a prime $^{\prime }$ denotes $\frac{d}{d\ell },$ defined by 
\begin{equation}
\frac{d}{d\ell }=\QATOPD\{ . {\left( \frac{d}{d\ell }\right) _{-}=\left(
r-r_{\circ }\right) \frac{d}{dr}}{\left( \frac{d}{d\ell }\right) _{+}=\sqrt{%
f\left( r\right) }\frac{d}{dr}}.
\end{equation}%
We observe that the $S_{0}^{0}$ component, proportional to the proper mass,
vanishes, i.e., $S_{0}^{0}=0.$ This can also be checked from the continuity
of the general mass formula%
\begin{equation}
m\left( r\right) \equiv \frac{r}{2}\left( 1-\left( \nabla r\right)
^{2}\right)
\end{equation}%
which gives 
\begin{equation}
m_{-}=m\left( r_{\circ }-0\right) =m_{+}=m\left( r_{\circ }+0\right) =\frac{%
r_{\circ }}{2}.
\end{equation}%
The surface pressures on the other hand become 
\begin{equation}
8\pi S_{2}^{2}=8\pi S_{3}^{3}=\frac{1}{r_{\circ }}\frac{d}{dr}\left( r\sqrt{%
f\left( r\right) }\right) -1.
\end{equation}%
In order to evaluate this expression we need to expand $f\left( r\right) $
in powers of $\left( r-r_{\circ }\right) .$ A detailed expansion process
gives%
\begin{eqnarray}
f\left( r\right) &=&\left( r-r_{\circ }\right) ^{2}-\frac{2}{3}\frac{\left(
r-r_{\circ }\right) ^{3}}{r_{\circ }}+\frac{1-2\ln 2}{3\ln 2}\frac{\left(
r-r_{\circ }\right) ^{4}}{r_{\circ }^{4}}-  \notag \\
&&\frac{1-10\ln 2}{15\ln 2}\frac{\left( r-r_{\circ }\right) ^{5}}{r_{\circ
}^{5}}+\frac{7-60\ln 2}{90\ln 2}\frac{\left( r-r_{\circ }\right) ^{6}}{%
r_{\circ }^{6}}+...
\end{eqnarray}%
From this expression, as the terms suggest, we can retain the quadratic term
as the leading order so that 
\begin{equation}
f\left( r\right) \tilde{=}\left( r-r_{\circ }\right) ^{2}.
\end{equation}%
Substituting this into (38) for the surface pressures, we obtain under (28)
that 
\begin{equation}
8\pi S_{2}^{2}=8\pi S_{3}^{3}=0.
\end{equation}%
At this point, it is instructive to calculate the charge to mass ratio for
such a particle (i.e., a black hole). In SI units we have 
\begin{equation}
\left( \frac{q}{m}\right) _{SI}=4\pi \sqrt{2G\epsilon _{\circ }}\left( \frac{%
q}{m}\right) _{geom.}=8\pi \sqrt{\frac{2G\epsilon _{\circ }}{\ln 2}}%
=1.04\times 10^{-9}\frac{C}{Kg}
\end{equation}%
which, predictably has a huge gap from the value of an electron $\left( \sim
1.7\times 10^{11}\frac{C}{Kg}\right) .$

Finally, we invoke the null-shell formalism \cite{13,14}, where the metrics
are cast into Kruskal form, 
\begin{equation}
ds^{2}\tilde{=}-F\left( u,v\right) dudv+r^{2}d\Omega ^{2}.
\end{equation}%
Here $F\left( u,v\right) $ is a bounded function on the horizon, and the
null coordinates are defined by 
\begin{eqnarray}
t-r_{\star } &=&u, \\
t+r_{\star } &=&v  \notag
\end{eqnarray}%
for $r_{\star }=\int \frac{dr}{f\left( r\right) }.$ By employing the
expansion (39) once more and, adopting its first term, we obtain the null
coordinates. The smooth matching on $u=0$ requires that \cite{14}%
\begin{equation}
\left( \frac{\partial r}{\partial u}\right) _{+}=\left( \frac{\partial r}{%
\partial u}\right) _{-}
\end{equation}%
implying in our case that it is satisfied for $r_{\circ }^{2}=\frac{1}{\ln 2}%
-1,$ which is nothing but the condition (28) that renders smooth matching
possible.

\section{Conclusion}

Employing a modified version of the BI action, consisting of nonpolynomial,
logarithmic parts, we obtain a class of regular, electrically-charged black
holes in Einstein-NED theory, which were previously unknown\cite{4}. Other
choices of boundary conditions, which we have not taken into consideration
in this paper, may give rise to what are called quasi-black holes (QBH). The
particular choice of the action provides a particle-like structure in flat
spacetime whose electric charge resides on its surface, while the particle
radius provides a natural cut-off for the electric field. This includes the
case of a massless particle whose entire mass derives from the electric
field energy. A similar picture applies to the curved space as well.
Remarkably, we uncover a regular class of purely electrically-charged black
hole solutions where for $r<r_{\circ }$, we have a uniform electric field
with $S_{\mu }^{\nu }=0$ at $r=r_{\circ }$. This class consists of the
extremal black hole in which the horizon, Born-Infeld parameter and charge
are related. Smooth gluing of a BR core to an outside RN was also known in
the Maxwell electrodynamics\cite{7}. The novel feature here is that the
horizon coincides with the specific value $r_{\circ }=\sqrt{\frac{1}{\ln 2}-1%
}$. This gives in SI units, $q_{SI}=1.50\times 10^{-18}C$ and $%
m_{SI}=1.44\times 10^{-9}kg$ for such a black hole.

\begin{acknowledgement}
We thank the anonymous referee for valuable and constructive suggestions.
\end{acknowledgement}

\bigskip

\bigskip

\bigskip

\bigskip

\end{document}